# Adaptive Basis Selection for Exponential Family Smoothing Splines with Application in Joint Modeling of Multiple Sequencing Samples

Ping Ma, Nan Zhang, Jianhua Z. Huang and Wenxuan Zhong [*]

University of Georgia, Fudan University and Texas A&M University

## Abstract

Second-generation sequencing technologies have replaced array-based technologies and become the default method for genomics and epigenomics analysis. Second-generation sequencing technologies sequence tens of millions of DNA/cDNA fragments in parallel. After the resulting sequences (short reads) are mapped to the genome, one gets a sequence of short read counts along the genome. Effective extraction of signals in these short read counts is the key to the success of sequencing technologies. Nonparametric methods, in particular smoothing splines, have been used extensively for modeling and processing single sequencing samples. However, nonparametric joint modeling of multiple second-generation sequencing samples is still lacking due to computational cost. In this article, we develop an adaptive basis selection method for efficient computation of exponential family smoothing splines for modeling multiple second-generation sequencing samples. Our adaptive basis selection gives a sparse approximation of smoothing splines, yielding a lower-dimensional effective model space for a more scalable computation. The asymptotic analysis shows that the effective model space is rich enough to retain essential features of the data. Moreover, exponential family smoothing spline models computed via adaptive basis selection are shown

[*]Ma's work was partially supported by NSF grant DMS-1438957, DMS-1440037, and NIH 1R01GM122080-01. Nan Zhang would like to acknowledge the support of Texas A&M Engineering Experiment Station and Texas A&M University Division of Research's 2014 Strategic Initiatives Seed Grant. Huang's work was partially supported by NSF grant DMS-1208952. Zhong's work was partially supported by NSF grant DMS-1440038, DMS-1406843, and NIH grant R01 GM113242-02. The authors thank Schmitz lab for providing methylation data and Lexiang Ji for explaining the methylation results. Ma and Zhang are equal contribution first authors

to have good statistical properties, e.g., convergence at the same rate as that of full basis exponential family smoothing splines. The empirical performance is demonstrated through simulation studies and two second-generation sequencing data examples.

# 1 Introduction

With the rapid development of biotechnologies, second-generation sequencing technologies have become default methods for various genomics and epigenomics analysis: RNA-seq for gene expression analysis (Mortazavi et al. (2008); Wilhelm et al. (2008); Nagalakshmi et al. (2008)), bisulfite sequencing for DNA methylation analysis (Cokus et al. (2008); Lister et al. (2008)), and ChIP-seq for genome-wide protein-DNA interaction analysis (Boyer et al. (2005); Johnson et al. (2007); Dixon et al. (2012)). Compared to their hybridization-based counterparts, e.g., microarry and ChIP-chip, second generation sequencing technologies offer up to a single-nucleotide resolution signal. Moreover, tens of millions of DNA or cDNA fragments can be sequenced in parallel. As the second-generation sequencing technologies become mature and cost-effective, conducting experiments with samples at multiple conditions, and/or of multiple tissue types, and/or at different time points becomes common. The large volume of data not only facilitates discovery in biology but also requires development of novel statistical methods for analysis.

After mapping the resulting sequences to a reference genome, researchers get a sequence of read counts, each of which corresponds to one nucleotide position, standing for the number of reads mapped onto that position. These short read counts may reflect certain biological interests, and statistical modeling and inference are indispensable for making discoveries (Li et al. (2010); Ji et al. (2014)). Whereas generalized linear models have been used in many studies (Li et al. (2010); Kuan et al. (2011); Dalpiaz et al. (2013)), versatile nonparametric modeling provides satisfactory performance (Zheng et al. (2011); Jaffe et al. (2012)). However, while each sequencing sample provides a genome size data set, multiple samples give rise to data sets of size in the tens of millions. Such a large volume of data makes application of many statistical models computationally infeasible. In this paper, we propose a scalable computational method for a class of flexible nonparametric regression models, exponential family smoothing splines (O'sullivan et al. (1986); Wahba et al. (1995)).

We first describe two typical contexts which motivate our methodology. Here and in the following, let $Y_i$ denote the $i$th read count, which associates with some covariates $x_i$, where $i = 1, \ldots, n$.

**Example 1.** *Profiling time course gene expression and isoform expression in RNA-Seq.* In many studies, researchers are interested in measuring the quantities of mRNAs molecules, i.e., quantifying gene/isoform expressions. mRNAs over time points in a certain biological process are quantified using RNA-seq. After mapping, read counts at each nucleotide position of the whole genome are obtained at each time point. A simple proposal for estimating gene/isoform expression is to average the short-read counts across all nucleotides within exons in each gene then normalize them by the total read counts (Cloonan et al. (2008)). However, the resulting gene/isoform expression levels may not be accurate due to significant sequencing bias of short-read counts (Dohm et al. (2008)). It has been observed that short-read counts at a nucleotide position tend to correlate with GC content (the percentage of bases that are either guanine or cytosine in a DNA sequence) in its neighborhood (Risso et al. (2011)). It is thus crucial to take into account the GC bias inherited in the RNA-seq technology while modeling the variation of short-read counts within each single gene over time. It is typical to use a regression model to remove GC bias. In particular, the response $Y_i$ is the short-read count of the $i$th nucleotide in a gene, covariates include time factor $t$ and a multivariate $x_i = (x_{\langle i1 \rangle}, \ldots, x_{\langle iK \rangle})$ which quantifies the GC content in the surrounding $K$ neighborhoods of $i$th nucleotide. Any model component involving $x_i$ is the GC bias and should be removed. □

**Example 2.** *Identifying differentially methylated regions using bisulfite sequencing.* DNA methylation is an essential epigenetic mechanism that regulates gene expression, cell differentiation, and development. A current technique for measuring DNA methylation levels is bisulfite sequencing. In our example, the methylation levels are measured at two different conditions. The goal is to compare the DNA methylation levels and identify the differentially methylated regions (DMRs). The total number of the mapped reads at the $i$th position is denoted as $N_i$, and that of methylated reads is denoted as $Y_i$. To identify the differentially methylated regions, we build a regression model between $(N_i, Y_i)$ and a bivariate covariate $x_i = (x_{\langle i1 \rangle}, x_{\langle i2 \rangle})$ where $x_{\langle i1 \rangle}$ relates to the genomic location and $x_{\langle i2 \rangle}$ is a condition indicator. We then detect the differential regions according to the diagnosis of the model components involving the condition indicator $x_{\langle i2 \rangle}$. □

Since short read data are clearly non-Gaussian, we adopt the exponential family of distributions, a rich family of distributions for non-Gaussian data. Specifically, the conditional distribution of $Y_i$ given the covariate $x_i$ is to have a density of the form

$$f(Y_i|x_i) = \exp[\{Y_i\eta(x_i) - b(\eta(x_i))\}/a(\phi) + c(Y_i, \phi)], \tag{1.1}$$

where $i = 1, \ldots, n$, $a > 0$, $b$ and $c$ are known functions, $\eta(x)$ is the regression function to be estimated, and $\phi$ is the dispersion parameter, assumed to be a constant, either known or considered as a nuisance parameter. The exponential family includes normal, binomial, Poisson, negative binomial, and many other distributions in a unified framework, and is broad enough to cover most practical applications in second-generation sequencing data.

Exponential family smoothing splines were developed by O'sullivan et al. (1986) for the univariate case and Wahba et al. (1995) extended them to the multivariate case. Wide application of exponential family smoothing splines has been hindered because of computational cost $O(n^3)$ where $n$ is sample size. Since the sample size of the second-generation sequencing data is in tens of millions, such computation is prohibitively expensive. Numerous solutions have been proposed to address the computational issue for smoothing splines with Gaussian responses (Luo and Wahba (1997); Kim and Gu (2004); Ma et al. (2015)). For exponential family smoothing splines, Gu and Kim (2002) obtained a lower-dimensional approximation of the estimates by randomly selecting a subset of basis functions. Such an approximation approach was also adopted by Kim and Gu (2004) for Gaussian regression. Ma et al. (2015) pointed that the simple random sampling strategy may fail to detect subtle signals when responses are Gaussian.

Extending the simple random sampling strategy of Gu and Kim (2002), we develop an adaptive basis selection method to construct a lower-dimensional space, called the effective model space, and then approximate exponential family smoothing splines there. Our basis sampling method uses the information from the response variable and thus distinguishes itself from the simple random basis sampling approach. A key adaptive step in our method is to slice the range of response variable. However, for second- generation sequencing data sets, different distribution assumptions require different slicing procedures. We provides practical guidance according to canonical parameter in data examples. As with the simple random sampling strategy of Gu and Kim (2002), the proposed method gives rise to a more scalable computation when approximating exponential family smoothing splines with large data sets. Because the response information is used in the sampling of basis functions, the adaptive basis selection provides more accurate estimates than the simple random sampling, as is evident in our simulation studies. Our asymptotic functional eigenvalue analysis shows the effective model space is rich enough to retain the essential information of true regression functions, and the approximated exponential family smoothing splines via adaptive basis selection converge to the truth at the same convergence rate as the full-basis exponential family smoothing splines. Our method is non-standard because of the response-dependent sampling scheme, and we provide practical guidelines for choosing the dimension of the effective model space.

The remainder of the article is organized as follows. In Section 2, we develop the adaptive basis selection method for exponential family smoothing splines. The asymptotic analysis is presented in Section 3. Simulation and data analysis follow in Sections 4 and 5. A few remarks in Section 6 conclude the article. Proofs of the theorems are in the supplementary material.

## 2 Efficient Computation via Adaptive Basis Selection

In this section, we review the penalized likelihood method for fitting exponential family smoothing spline models and investigate the computation complexity, then develop the adaptive basis selection method to efficiently approximate the estimator in a low-dimensional function space.

### 2.1 Penalized likelihood approach

We estimate $\eta$ by minimizing the penalized likelihood functional

$$-\frac{1}{n}\sum_{i=1}^{n}\{Y_i\eta(x_i) - b(\eta(x_i))\} + \frac{\lambda}{2}J(\eta), \tag{2.1}$$

where the first term is derived from the negative log likelihood, and $J(\eta) = J(\eta, \eta)$ is a quadratic functional penalizing the roughness of $\eta$. The smoothing parameter $\lambda$ then controls the trade-off between the goodness-of-fit and smoothness of $\eta$.

**Example 1 (continued)** *Profiling time course gene expressions in RNA-Seq.* We assume the short-read count $Y$ given the covariate $x$ is Poisson distributed, $Y|x \sim \text{Poisson}(\lambda(x))$ with density $\lambda(x)^Y e^{-\lambda(x)}/Y!$. Here $\eta(x) = \log \lambda(x)$ at (1.1). Sun et al. (2016) models the read count $Y$ by a negative binomial distribution to account for excessive variation in read counts. Thus, $Y|x \sim \text{NegBinomial}(r, p(x))$ with density $\binom{Y+r-1}{Y}p(x)^Y(1-p(x))^r$, so $\eta(x) = \log p(x)$. □

**Example 2 (continued)** *Identifying differentially methylated regions.* We assume the number of methylated reads $Y$ given covariate $x$ at position is binomial, $Y|x \sim \text{Binomial}(N, p(x))$ with density $\binom{N}{Y}p(x)^Y(1-p(x))^{N-Y}$ and $\eta(x) = \log\{p(x)/(1-p(x))\}$. □

The standard formulation of smoothing splines restricts minimizing (2.1) to a reproducing kernel Hilbert space (RKHS) $\mathcal{H} = \{\eta : J(\eta) < \infty\}$. To prevent interpolation, the null space of $J$, $\mathcal{N}_J = \{\eta : J(\eta) = 0\}$, is assumed to be a finite dimensional linear subspace of $\mathcal{H}$ with basis $\{\phi_i: i = 1, \ldots, m\}$. Denote the orthogonal decomposition of $\mathcal{H}$ by $\mathcal{N}_J \oplus \mathcal{H}_J$ where $\mathcal{H}_J$ is still a reproducing kernel Hilbert space. Let $R_J(x, y)$ be the reproducing kernel of $\mathcal{H}_J$. The representer theorem (Wahba (1990)) shows that the minimizer of (2.1) in the RKHS $\mathcal{H}$ has the simple form

$$\eta(x) = \sum_{\nu=1}^{m} d_\nu \phi_\nu(x) + \sum_{i=1}^{n} c_i R_J(x_i, x), \tag{2.2}$$

where coefficients $d_\nu$ and $c_i$ are to be estimated from data.

When $x$ is multivariate, the functional analysis of variance (ANOVA) decomposition of a function $\eta$ is

$$\eta(x) = \eta_0 + \sum_{j=1}^{d} \eta_j(x_{\langle j \rangle}) + \sum_{j=1}^{d} \sum_{k=j+1}^{d} \eta_{jk}(x_{\langle j \rangle}, x_{\langle k \rangle}) + \cdots + \eta_{1,\ldots,d}(x_{\langle 1 \rangle}, \ldots, x_{\langle d \rangle}), \qquad (2.3)$$

where the $\eta_0$ is a constant, the $\eta_j$'s are the main effects, the $\eta_{jk}$'s are the two-way interactions, etc. The identifiability of the terms in (2.3) is ensured by side conditions through averaging operators (Wahba (1990); Gu (2013)). When estimating $\eta$ from (2.1) with structure (2.3), we consider $\eta_j \in \mathcal{H}_{\langle j \rangle}$, where $\mathcal{H}_{\langle j \rangle}$ is an RKHS with tensor sum decomposition $\mathcal{H}_{\langle j \rangle} = \mathcal{H}_{0\langle j \rangle} \oplus \mathcal{H}_{1\langle j \rangle}$, $\mathcal{H}_{0\langle j \rangle}$ is the finite-dimensional "parametric" subspace consisting of parametric functions, and $\mathcal{H}_{1\langle j \rangle}$ is the "nonparametric" subspace consisting of smooth functions. The induced tensor product space is

$$\mathcal{H} = \otimes_{j=1}^{d} \mathcal{H}_{\langle j \rangle} = \oplus_{\mathcal{S}} [(\otimes_{j \in \mathcal{S}} \mathcal{H}_{1\langle j \rangle}) \otimes (\otimes_{j \notin \mathcal{S}} \mathcal{H}_{0\langle j \rangle})] = \oplus_{\mathcal{S}} \mathcal{H}_{\mathcal{S}},$$

where the summation runs over all subsets $\mathcal{S} \subseteq \{1, \ldots, d\}$. The corresponding penalty function is $J(\eta) = \sum_{\mathcal{S}} \theta_{\mathcal{S}}^{-1} J_{\mathcal{S}}(\eta_{\mathcal{S}})$ with $\eta_{\mathcal{S}} \in \mathcal{H}_{\mathcal{S}}$, where $\theta_{\mathcal{S}} > 0$ are extra smoothing parameters, and $J_{\mathcal{S}}$ is the square norm in $\mathcal{H}_{\mathcal{S}}$. The subspaces $\mathcal{H}_{\mathcal{S}}$ form two large subspaces: $\mathcal{N}_J = \{\eta : J(\eta) = 0\}$, which is the null space of $J(\eta)$, and $\mathcal{H} \ominus \mathcal{N}_J$ with the reproducing kernel $R_J = \sum_{\mathcal{S}} \theta_{\mathcal{S}} R_{\mathcal{S}}$ where $R_{\mathcal{S}}$ is the reproducing kernel in $\mathcal{H}_{\mathcal{S}}$. The smoothing spline estimator in such a reproducing kernel Hilbert space is called a tensor product smoothing spline.

In Example 2, covariates are of mixed types, continuous and discrete. Consider a bivariate function $\eta(x, \tau)$, where $x \in [0, 1]$ and $\tau \in \{1, \ldots, t\}$. One can write $\eta(x, \tau) = \eta_\emptyset + \eta_1(x) + \eta_2(\tau) + \eta_{1,2}(x, \tau)$, where $\eta_\emptyset$ is a constant, $\eta_1(x)$ is a function of $x$ satisfying $\eta_1(0) = 0$, $\eta_2(\tau)$ is a function of $\tau$ satisfying $\sum_{\tau=1}^{t} \eta_2(\tau) = 0$, and $\eta_{1,2}(x, \tau)$ satisfies $\eta_{1,2}(0, \tau) = 0$, $\forall \tau$, with $\sum_{\tau=1}^{t} \eta_{1,2}(x, \tau) = 0$, $\forall x$. Regarding the quadratic functional $J$, one can use

$$J(\eta) = \theta_1^{-1} \int_0^1 (d^2\eta_1/dx^2)^2 dx + \theta_{1,2}^{-1} \int_0^1 \sum_{\tau=1}^{t} (d^2\eta_{1,2}/dx^2)^2 dx.$$

The null space $\mathcal{N}_J$ has dimension $2t$ with basis functions

$$\{1, x, I_{[\tau=j]} - 1/t, (I_{[\tau=j]} - 1/t)x, j = 1, \ldots, t-1\}.$$

The reproducing kernel of $\mathcal{H}_J$ is

$$R_J(x_1, \tau_1; x_2, \tau_2) = \theta_1 \int_0^a (x_1 - u)_+(x_2 - u)_+ du + \theta_{1,2}(I_{[\tau_1=\tau_2]} - 1/t) \int_0^a (x_1 - u)_+(x_2 - u)_+ du.$$

General discussions of reproducing kernels can be found in Section 2.4 of Gu (2013).

For the exponential family, $E[Y|x] = b'(\eta(x)) = \mu(x)$ and $\text{var}[Y|x] = b''(\eta(x))a(\phi) = \nu(x)a(\phi)$. When the likelihood function at (2.1) has a unique minimizer in $\mathcal{N}_J$, the minimizer $\hat{\eta}$ of (2.1) uniquely exists. Fixing the smoothing parameter $\lambda$ (and ones hidden in $J(\eta)$, if present), (2.1) may be minimized through a Newton iteration. Write $l(\eta(x_i); Y_i) = -Y_i\eta(x_i) + b(\eta(x_i))$, $u(\eta(x_i); Y_i) = -Y_i + b'(\eta(x_i))$, and $w(\eta(x_i); Y_i) = b''(\eta(x_i)) = \nu(x_i)$. The quadratic approximation of $l(\eta(x_i); Y_i)$ at the current estimate $\tilde{\eta}(x_i)$ is given by

$$l(\eta(x_i; Y_i) \approx l(\tilde{\eta}(x_i); Y_i) + \tilde{u}_i(\eta(x_i) - \tilde{\eta}(x_i)) + \tilde{w}_i(\eta(x_i) - \tilde{\eta}(x_i))^2/2 = \tilde{w}_i(\tilde{Y}_i - \eta(x_i))^2/2 + C_i,$$

where $\tilde{u}_i = u(\tilde{\eta}(x_i); Y_i)$, $\tilde{w}_i = w(\tilde{\eta}(x_i); Y_i)$, $\tilde{Y}_i = \tilde{\eta}(x_i) - \tilde{u}_i/\tilde{w}_i$ and $C_i$ is independent of $\eta(x_i)$. The Newton iteration can thus be performed by minimizing the penalized weighted least squares,

$$\sum_{i=1}^{n} \tilde{w}_i(\tilde{Y}_i - \eta(x_i))^2 + n\lambda J(\eta). \tag{2.4}$$

Although fast algorithms (Reinsch (1967)) are available when $x$ is univariate, solving the problem for multivariate $x$ requires $O(n^3)$ operations, see Section 3.4 of Gu (2013). The high computational cost of smoothing splines renders it inapplicable for modeling second-generation sequencing data. In our two examples, sample sizes are 48,660 and 22,588.

## 2.2 Adaptive basis selection

To alleviate the computational cost of smoothing splines, one can restrict the minimizer of (2.1), or equivalently (2.4), to a reduced subspace of $\mathcal{H}$. Such a subspace is called an effective model space. Following Ma et al. (2015), we develop an adaptive basis sampling approach to selecting a subset of full basis functions and constructing an effective model space. For an effective model space the computational cost in constructing it is cheap, and the essential information of the true function $\eta$ is retained.

**Adaptive basis selection algorithm.**

(1) Divide the range of the responses $\{Y_i\}_{i=1}^{n}$ or derived data into $K$ disjoint intervals, denoted by $S_1, S_2, \ldots, S_K$.

(2) For $k = 1, \ldots, K$, take a random sample $x_1^{*(k)}, \ldots, x_{n_k}^{*(k)}$ of size $n_k$ with replacement, from original sample $x_i$ with probability $|S_k|^{-1} I_{y_i \in S_k}$, where $|S_k|$ is the number of observations in $S_k$. Denote the combined sample as $x_1^*, \ldots, x_{n^*}^*$ with sample size $n^*$.

(3) Minimize (2.1) over

$$\mathcal{H}_E = \mathcal{N}_J \oplus \text{span}\{R_J(x_j^*, \cdot), j = 1, \ldots, n^*\}$$

where $\mathcal{H}_E$ is the effective model space. The minimizer has the expression

$$\hat{\eta}_A(x) = \sum_{i=1}^{m} d_\nu \phi_\nu(x) + \sum_{j=1}^{n^*} c_j R_J(x_j^*, x) \qquad (2.5)$$

where $\hat{\eta}_A(x)$ is an exponential family smoothing spline estimate through adaptive basis selection.

When dividing the range of response variable at step (1), we take the specific exponential family distribution assumption into account. In Example 1, responses follow a Poisson distribution and we can apply slicing directly on $\{Y_i\}_{i=1}^n$. In Example 2, $Y_i$ is binomial, we propose to divide the range of ratio $Y_i/N_i$ to avoid possible heterogeneity in count data. Such ratios are empirical estimates of the success probabilities that are connected with the canonical parameter of binomial distribution monotonically.

Substituting (2.5) into (2.4), the numerical problem is to minimize

$$(\tilde{\mathbf{Y}} - S\mathbf{d} - R\mathbf{c})^T \tilde{W} (\tilde{\mathbf{Y}} - S\mathbf{d} - R\mathbf{c}) + n\lambda \mathbf{c}^T Q \mathbf{c} \qquad (2.6)$$

with respect to $\mathbf{d}$, $\mathbf{c}$, where $\tilde{\mathbf{Y}} = (\tilde{Y}_1, \ldots, \tilde{Y}_n)^T$, $S$ is $n \times m$ with the $(i,\nu)$th entry $\phi_\nu(x_i)$, $R$ is $n \times n^*$ with the $(i,j)$th entry $R_J(x_i, x_j^*)$, $Q$ is $n^* \times n^*$ with the $(j,k)$th entry $R_J(x_j^*, x_k^*)$, and $\tilde{W} = \text{diag}(\tilde{w}_1, \ldots, \tilde{w}_n)$.

The tuning parameter $\lambda$ (including $\theta$) is chosen by generalized approximate cross-validation (Gu and Xiang (2001)). See more details in the supplementary material.

## 3 Asymptotic Analysis

We develop an asymptotic analysis analogous to that in Ma et al. (2015) to guide the construction of the effective model space and establish the convergence rate of the smoothing spline with adaptive basis selection. Proofs of results in this section are in the supplementary materials.

### 3.1 Regularity conditions and rate of convergence

We have $l(\eta(x); y) = -y\eta(x) + b(\eta(x))$ and $u(\eta; y) = dl/d\eta$, $w(\eta; y) = d^2l/d\eta^2$. Assume that

$$\text{E}\{u(\eta_0(X); Y)|X\} = 0, \quad \text{E}\{u^2(\eta_0(X); Y)|X\} = \sigma^2 \text{E}\{w(\eta_0(X); Y)|X\},$$

and write $v_\eta(x) = \text{E}\{w(\eta(x); Y)|X = x\}$. Let $f_X(\cdot)$ be the marginal density of the predictor variable $X$ and take

$$V(g) = \int_{\mathcal{X}} g^2(x) v_{\eta_0}(x) f_X(x) \, dx.$$

**Condition 1.** $V$ is completely continuous with respect to $J$.

Here, there exists a sequence of eigenfunctions $\phi_\nu \in \mathcal{H}$ and the associated nonnegative increasing sequence of eigenvalues $\rho_\nu$ such that $V(\phi_\nu, \phi_\mu) = \delta_{\nu\mu}$ and $J(\phi_\nu, \phi_\mu) = \rho_\nu \delta_{\nu\mu}$ where $\delta_{\nu\mu}$ is the Kronecker delta.

**Condition 2.** For some $r > 1$ and $\beta > 0$, we have $\rho_\nu > \beta \nu^r$ for sufficiently large $\nu$.

The growth rate of the eigenvalues $\rho_\nu$ of $J$ with respect to $V$ essentially dictates how fast $\lambda$ should approach to zero. See Section 9.1 of Gu (2013).

**Condition 3.** For $\eta$ in a convex set $B_0$ around $\eta_0$ containing $\hat{\eta}$ and $\tilde{\eta}$,

$$c_1 \, w(\eta_0(x); y) \leqslant w(\eta(x); y) \leqslant c_2 \, w(\eta_0(x); y)$$

holds uniformly for some $0 < c_1 < c_2 < \infty$, $\forall x \in \mathcal{X}$, $\forall y$.

Roughly speaking, Condition 3 concerns the equivalence of information within $B_0$.

**Condition 4.** There is a constant $c_3 < \infty$ such that $\operatorname{var}\{\phi_\nu(X)\phi_\mu(X)w(\eta_0(X);Y)\} \leqslant c_3$ for all $\nu, \mu$.

As the $\phi_\nu$'s forms an orthonormal system relative to $V(\cdot, \cdot)$ such that

$$\mathrm{E}\{\phi_\nu(X)\phi_\mu(X)w(\eta_0(X);Y)\} = V(\phi_\nu, \phi_\mu) = \delta_{\nu\mu},$$

one has $\mathrm{E}\{\phi_\nu^2(X)\phi_\mu^2(X)w^2(\eta_0(X);Y)\} \leqslant c_3 + 1$. Condition 4 basically requires that the fourth moments of the $\phi_\nu(X)$ be uniformly bounded.

**Theorem 1.** *If $\sum_i \rho_i^p V(\eta_0, \phi_i)^2 < \infty$ for some $p \in [1, 2]$, and Conditions 1-4 hold, as $\lambda \to 0$ and $n^* \lambda^{2/r} \to \infty$, we have*

$$(V + \lambda J)(\hat{\eta}_A - \eta_0) = O_p(n^{-1}\lambda^{-1/r} + \lambda^p).$$

*In particular, when $\lambda \asymp n^{-r/(pr+1)}$, the estimator $\hat{\eta}_A$ achieves the optimal convergence rate*

$$(V + \lambda J)(\hat{\eta}_A - \eta_0) = O_p(n^{-pr/(pr+1)}).$$

Thus, under regularity conditions, the convergence rate of the smoothing spline estimator using an adaptively selected basis is the same as that of the smoothing spline estimator using the full basis, see Theorem 9.17 in Gu (2013).

## 3.2 The dimension of the effective model space

Utilizing our asymptotic analysis results, we can determine the dimension of the effective model space $\mathcal{H}_E$. On one hand, Theorem 1 requires $n^*\lambda^{2/r} \to \infty$. When $\lambda \asymp n^{-r/(pr+1)}$, we can choose $n^* \asymp n^{2/(pr+1)+\delta}$, where $\delta$ is an arbitrary small positive number. On the other hand, constant $p$ depends on the smoothness of $\eta$: for roughest $\eta$ satisfying $J(\eta) < \infty$, we have $p = 1$, whereas for the smoothest $\eta$, we have $p = 2$.

Take the univariate cubic smoothing spline as an example: $J(\eta) = \int (\eta'')^2$ with $r = 4$ and $\lambda \asymp n^{-4/(4p+1)}$. The proper dimension of the effective model space is $n^* = n^{2/(4p+1)} + \delta$, which is in the range of $O(n^{2/9+\delta})$ and $O(n^{2/5+\delta})$ for $p \in [1, 2]$. For the linear smoothing spline, $J(\eta) = \int (\eta')^2$ and $r = 2$. The dimension of the effective model space ranges from $O(n^{2/5+\delta})$ to $O(n^{2/3+\delta})$ for $p$ in $[1, 2]$. In our simulations and examples, we take dimension of the effective model space $n^*$ to be between $4n^{2/9}$ and $20n^{2/9}$ for the cubic smoothing spline with selected basis, and between $4n^{2/5}$ and $20n^{2/5}$ for the linear smoothing spline with selected basis.

## 4 Simulation Study

We approximated the exponential family smoothing spline estimate via adaptive basis sampling and that with uniform basis sampling (Kim and Gu (2004)) to three multivariate test functions. Exponential family distributions considered here include the negative binomial, Poisson, and binomial. When generating predictors $x$, a random design was adopted: $n = 1600$ points were uniformly generated from the domains. Responses were correspondingly generated under each distribution assumption. The number of slices was suggested by Scott's method (Scott (1992)) and based on our asymptotic results, the dimension of the effective model space was set to $10n^{2/9}$, which meant $n^* = 52$ basis functions were sampled for both sampling methods for approximating exponential family smoothing splines.

We first took the bivariate blocks function with negative binomial distribution as an example. The bivariate blocks function is a direct generalization of the univariate blocks function (Donoho and Johnstone (1994)) to bivariate case. Let blocks($\cdot$) be the univariate blocks function, then the bivariate blocks function is $\text{blocks}_2(x_{\langle 1 \rangle}, x_{\langle 2 \rangle}) = \text{blocks}(x_{\langle 1 \rangle})$. For the negative binomial distribution with parameters $(\alpha, p)$, we set the success probability at $p = (\text{blocks}_2 + 2.5)/8$ and the target number of successful trials at $\alpha = 3$.

Examples were constructed from the joint probability density of a $d$-dimensional nonparanormal distribution (Liu et al. (2009))

$$p_\alpha^d(x) = (2\pi)^{-d/2} |\Sigma|^{-1/2} \exp\left\{-\frac{1}{2} f(x)^\top \Sigma^{-1} f(x)\right\} \prod_{j=1}^d |f'_j(x_j)|, \qquad (4.1)$$

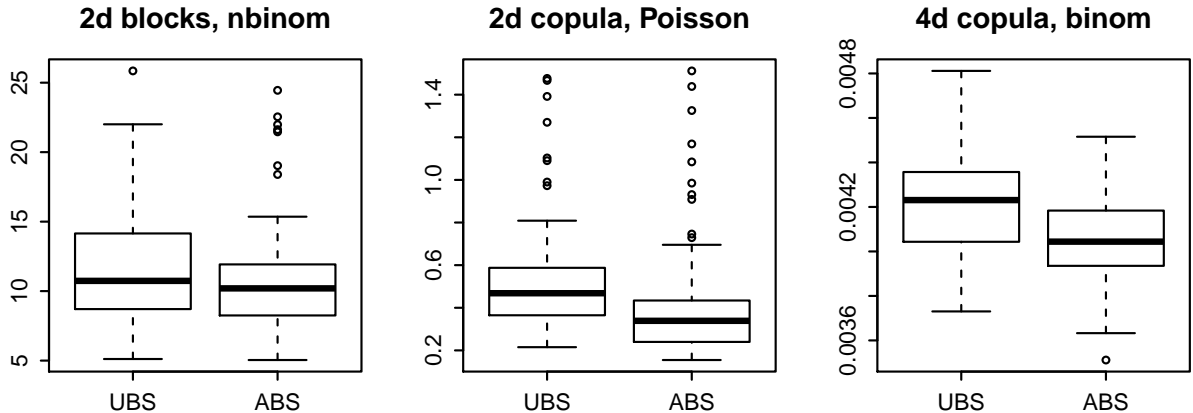

Figure 1: Boxplots of MSE for multivariate simulation studies. Left: bivariate blocks function with negative binomial distribution; middle: bivariate copula density function with Poisson distribution; right: four-dimensional copula density function with binomial distribution. UBS and ABS stand for exponential family smoothing spline estimator under uniform and adaptive basis sampling strategies.

where $\Sigma$ was a $d \times d$ matrix with diagonal entries 1, super and sub diagonal entries 0.5 and other entries 0; the $j$th component of $f(x)$ was $f_j(x) = \alpha_j \operatorname{sign}(x) |x|^{\alpha_j}$ with $\alpha_j$'s as shape parameters.

A second example was a bivariate copula density function with Poisson distribution. The bivariate copula density was obtained by setting $d = 2$ and $\alpha = (2, 3)^\top$ in (4.1). For the Poisson distribution, the mean parameter was $\lambda = 1 + 2(2\pi)^{p/2} |\Sigma|^{1/2} p_\alpha^d$. Our third example was a four-dimensional copula density function with binomial distribution. We took $d = 4$ and $\alpha = (0.1, 0.1, 0.1, 0.1)^\top$ in (4.1). For binomial distribution with parameters $(m, p)$, the number of trials was $m = 50$ and the success probability was $p = \exp(p_\alpha^d)/\{1 + \exp(p_\alpha^d)\}$.

To evaluate the performance of each approximation method, we repeated the experiment 100 times under each simulation set-up, and calculated the mean squared error (MSE) for the estimate. For the binomial and negative binomial distributions MSE $= \sum_{i=1}^{n} \{\hat{p}(x_i) - p(x_i)\}^2$, and for the Poisson distribution MSE $= \sum_{i=1}^{n} \{\hat{\lambda}(x_i) - \lambda(x_i)\}^2$. Boxplots of MSEs for the three multivariate test functions are displayed in Figure 1. The proposed adaptive basis sampling scheme enables exponential family smoothing splines to be more accurate and stable. Further calculation shows that, under the three simulation set-ups, smoothing splines with adaptive basis sampling outperform those with uniform basis sampling 69, 96 and 76 times of 100 experiments, respectively.

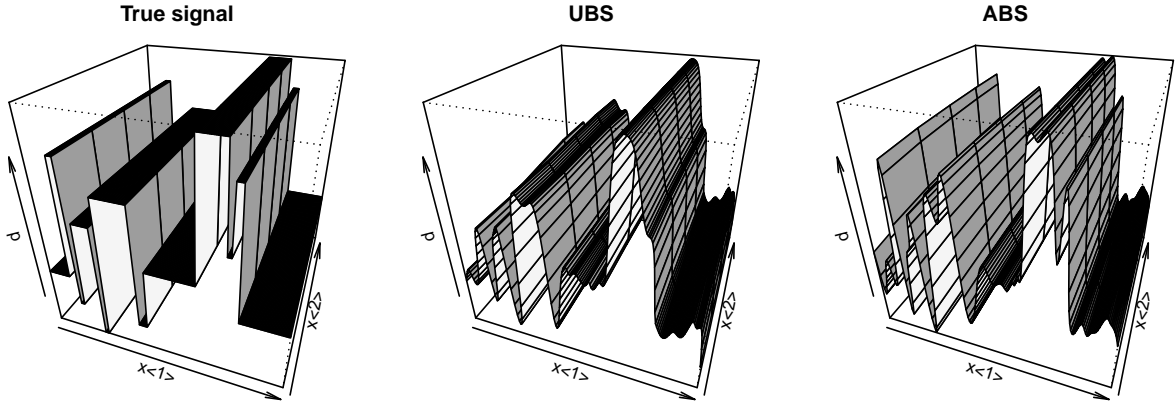

Figure 2: Bivariate blocks function with negative binomial distribution. Perspective plots of true probability, fitted values by smoothing splines via uniform basis sampling and adaptive basis sampling.

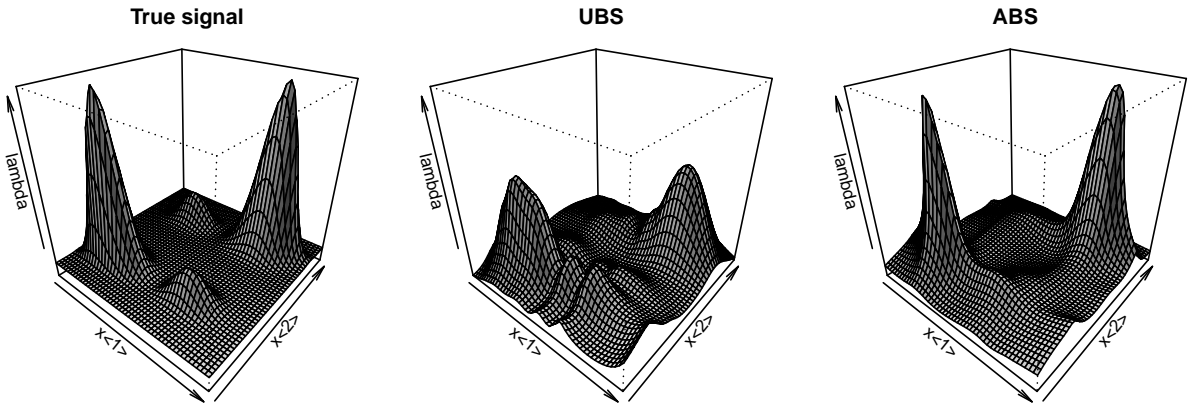

Figure 3: Bivariate copula density function with Poisson distribution. Perspective plots of true mean parameter, fitted values by smoothing splines via uniform basis sampling and adaptive basis sampling.

Figures 2 and 3 display the visualization for the two bivariate examples for a single run. In Figure 2, the probability parameter of the negative binomial distribution is a bivariate blocks function that has many abrupt local jumps in the $x_{\langle 1 \rangle}$ direction. The proposed method successfully recovers such fine scale information while the uniform basis sampling fails. In Figure 3, the mean parameter of the Poisson distribution behaves relatively smoothly. There are four peaks across the domain: two are significantly higher than the others. The estimate with adaptively sampled basis apparently provides a better fit: the two large peaks recovered are closer to the truth.

## 5 Examples

In this section, we analyze the data sets from Examples 1 and 2.

### 5.1 Modeling the time course gene expression and isoform expression profiles using RNA-Seq

*Drosophila melanogaster* (fruit fly) shares a substantial genetic content with humans and has been used as a translational model for human development. To study Drosophila melanogaster development, Graveley et al. (2011) conducted time course RNA-seq experiments. In these experiments, the authors collected 12 embryonic RNA samples at two-hour intervals for 24 hours in the stage of early embryos. The samples were then sequenced using an Illumina Genome Analyzer IIx platform.

We are interested in estimating time course gene and isoform expressions at the early embryos stage. It is necessary to take into account the sequencing bias, in particular the GC bias. We fulfill the task in two steps. First, we attempt a nonparametric model to model time course gene expression profiles while accounting for the GC bias. Since the read in Graveley et al. (2011) is 76 base-pair long, we count the GC content in each read length interval. We denote short-read counts at the $j$th nucleotide at time point $t$ by $Y_{jt}$, the number of GC counts in the neighborhood of 1 to 76 nucleotides away from the $j$th nucleotide by $x_{\langle 1j \rangle}$, that in the neighborhood of 77 to 152 nucleotides away from the $j$th nucleotide by $x_{\langle 2j \rangle}$, and that in the neighborhood of 153 to 228 nucleotides away from the $j$th nucleotide by $x_{\langle 3j \rangle}$. We built a nonparametric Poisson model for the short-read counts: $Y_{jt} \sim \text{Poisson}(\lambda_{jt})$, with

$$\log(\lambda_{jt}) = C + \eta_0(j,t) + \eta(x_{\langle 1j \rangle}, x_{\langle 2j \rangle}, x_{\langle 3j \rangle}), \tag{5.1}$$

where $\eta_0(j,t)$ represents the time trend along the gene, and $\eta(x_{\langle 1ij \rangle}, x_{\langle 2ij \rangle}, x_{\langle 3ij \rangle})$ is the sequencing bias due to GC content. We applied functional ANOVA decomposition to $\eta_0(j,t)$ and $\eta(x_{\langle 1 \rangle}, x_{\langle 2 \rangle}, x_{\langle 3 \rangle})$ and all main effects, two-way and three-way interactions of covariates were kept $\eta_0(j,t) = C_1 + \eta_{01}(j) + \eta_{02}(t) + \eta_{012}(j,t)$,

$$\eta(x_{\langle 1j \rangle}, x_{\langle 2j \rangle}, x_{\langle 3j \rangle}) = C_2 + \sum_{k=1}^{3} \eta_k(x_{\langle kj \rangle}) + \sum_{k=1}^{3} \sum_{l=j+1}^{3} \eta_{kl}(x_{\langle kj \rangle}, x_{\langle lj \rangle}) + \eta_{123}(x_{\langle 1j \rangle}, x_{\langle 2j \rangle}, x_{\langle 3j \rangle}).$$

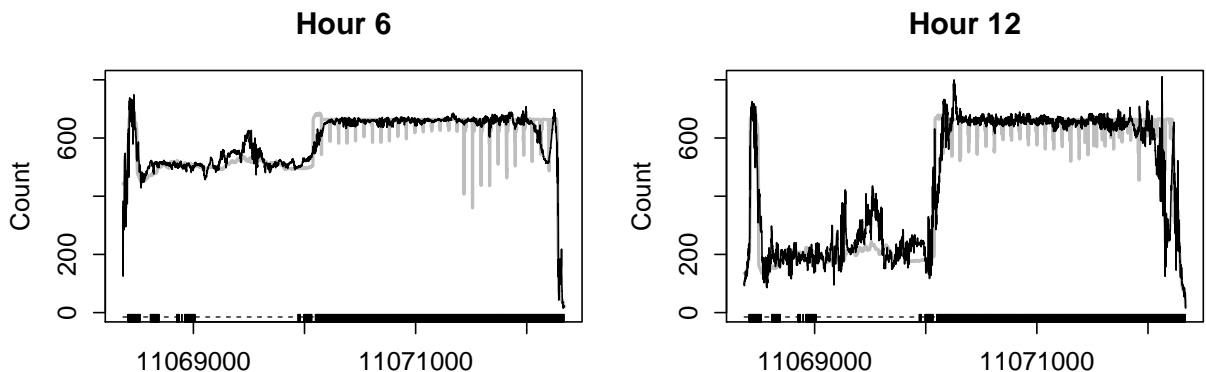

Figure 4: Estimated counts after removing GC bias for two time courses of gene Hsc70-4. Observed counts are in gray line and black line is the estimation, the blocks in the bottom are exons.

We then fit the exponential family smoothing spline models using the penalized likelihood (2.1) for each gene. The total number of observations during all twelve time courses is over $50,000$, which renders standard computations infeasible. Instead, we used the proposed adaptive basis sampling method when fitting the exponential family smoothing spline models.

Here is detailed algorithmic information of the analysis of the RNA-seq data of two genes: heat shock protein cognate 4 (Hsc70-4), and elongation factor 2b (Ef2b), which are $3,974$ and $4,055$ bp long when only exons are kept, respectively. Using our adaptive basis selection method for fitting exponential family smoothing spline models, we took the dimension of the effective model space $n^* = 72$ for both genes. The computing times for running the exponential family smoothing spline models with adaptive basis selection were 95 and 124 CPU seconds on a 2.90 GHz Intel Xeon computer. To assess the adequacy of the exponential family smoothing splines estimates via adaptive basis selection, we computed the quasi-$R^2$ (Li et al. (2010)),

$$R^2 = 1 - d/d_0 \tag{5.2}$$

where $d$ is the deviance of the fitted model and $d_0$ is the deviance of the null model with only constant mean. The quasi-$R^2$ of the fitted exponential family smoothing spline model via adaptive sampling method was 0.87 for Hsc70-4, and 0.86 for Ef2b. Figure 4 and 5 display the estimated counts $\exp\{\alpha_i + \eta_0(t)\}$ by removing GC bias $\eta$ from $\lambda_{ijt}$ in two genes.

As a second step, we estimated the isoform expression using the Poisson model and maximum likelihood method developed in Jiang and Wong (2009). Consider a gene with $m$ exons of lengths $(l_1, l_2, ..., l_m)$ and $k$ isoforms with expressions $\boldsymbol{\theta} = (\theta_1, \theta_2, ..., \theta_k)$. Suppose $z_i$ be GC biased-corrected read counts (obtained in the first step) fall in $i$th exon. We assumed that $z_i$ is Poisson with mean $\lambda_i$. For instance, the $\lambda_i$ for the number of reads falling into exon $i$ is $l_i \sum_{j=1}^{k} c_{ij} \theta_j$, where $c_{ij}$ is an indicator function: 1 if isoform $j$ contains exon $i$, and 0 otherwise. Then a Poisson model was fitted to $z_i$ using maximum likelihood approach, see details in Jiang and Wong (2009). According to flybase (www.flybase.org) annotation, there are seven known isoforms for Hsc70-4 gene and three for Ef2b. We estimated the isoform expression for both Hsc70-4 and Ef2b. The estimated isoform expressions at Hour 6 and 12 for Hsc70-4 and those for Ef2b at Hour 14 and 20 are listed in Table 1 and 2.

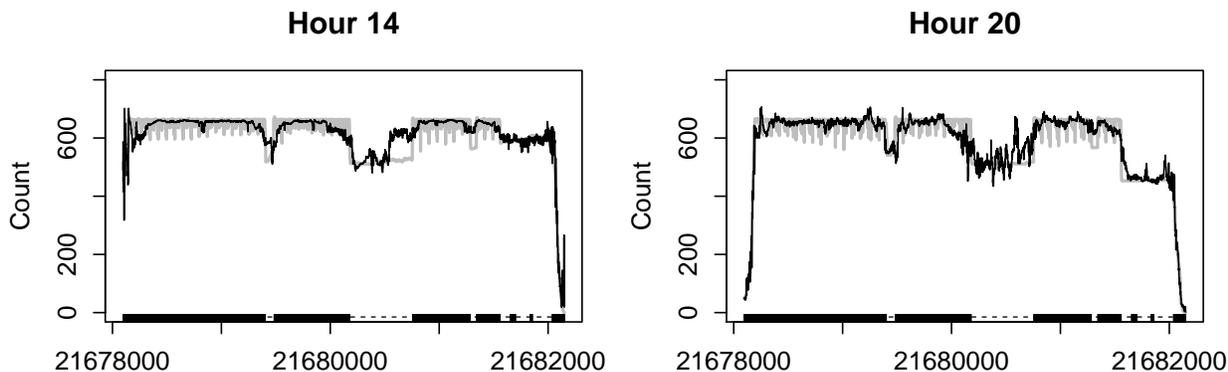

Figure 5: Predicted counts after removing GC bias for two time courses of gene Ef2b. Observed counts are in gray line and black line is the estimation, the blocks in the bottom are exons.

Table 1: Raw read counts and fitted counts for all seven isoforms of gene Hsc70-4 at Hour 6 and 12.

|  | Hour 6 | | | | | | |
| --- | --- | --- | --- | --- | --- | --- | --- |
|  | Isoform 1 | Isoform 2 | Isoform 3 | Isoform 4 | Isoform 5 | Isoform 6 | Isoform 7 |
| Raw | 1522416 | 1492814 | 1466414 | 1468447 | 1503038 | 1495502 | 1506437 |
| Fitted | 1503707 | 1474983 | 1446416 | 1448412 | 1482390 | 1476911 | 1488867 |
|  | Hour 12 | | | | | | |
|  | Isoform 1 | Isoform 2 | Isoform 3 | Isoform 4 | Isoform 5 | Isoform 6 | Isoform 7 |
| Raw | 1486773 | 1443093 | 1435028 | 1435856 | 1450492 | 1445258 | 1450323 |
| Fitted | 1441400 | 1399584 | 1388391 | 1389162 | 1401834 | 1401303 | 1408558 |

## 5.2 Differentially methylated DNA regions in *Arabidopsis*

DNA methylation is an important epigenetic mechanism that regulates gene expression, cell differentiation, and development. The whole genome GC methylation levels of four strains of *Arabidopsis thaliana* were measured using whole genome bisulfite sequencing (Ji et al., 2014). The whole genome of Arabidopsis is around 135 million bp. Two strains were from one generation and the other two were taken from a second generation. The total number of GC methylated nucleotides is 23,361.

Table 2: Raw read counts and fitted counts for all three isoforms of gene Ef2b at Hour 14 and 20.

|        | Hour 14 | | | Hour 20 | | |
|--------|---------|---------|---------|---------|---------|---------|
|        | Isoform 1 | Isoform 2 | Isoform 3 | Isoform 1 | Isoform 2 | Isoform 3 |
| Raw    | 1824904 | 1809689 | 1824718 | 1773156 | 1766892 | 1778588 |
| Fitted | 1798631 | 1781373 | 1796776 | 1749174 | 1742474 | 1754925 |

Let $Y_{i,s,g}$ and $N_{i,s,g}$ be the methylated and total read counts at genetic position $i$ in strain $s$ of generation $g$, where $s = 1, 2$ and $g = 1, 2$. We built a nonparametric mixed-effect binomial model (Gu and Ma (2005)) for the short-read counts, $Y_{i,s,g} \sim \text{Binomial}(N_{i,s,g}, p(i,s,g))$, with the canonical parameter specified by a nonparametric component and a random effect,

$$\log \frac{p(i,s,g)}{1-p(i,s,g)} = \eta(i,g) + b_s,$$

where $\eta$ has a functional ANOVA structure,

$$\eta(i,g) = \eta_C + \eta_1(i) + \eta_2(g) + \eta_{12}(i,g), \tag{5.3}$$

and the random effect $b_s \sim N(0, \sigma^2)$ induces the strain correlation.

Our primary goal is to identify the differentially methylated regions between two generations. Since the genome of *Abrabidopsis* is around 135 million bp, it is computational impossible to apply smoothing spline models directly to the whole genome. (Gu and Ma (2005)) suggested using the simple random basis sampling method of Gu and Kim (2002) to reduce computational cost for large sample data. Instead, we use the adaptive basis selection. We first divided the whole genome into segments of length 20 k bp and then did an exhaustive search among the segments. We fit exponential family smoothing splines to each segment using the adaptive basis selection method, and employed an inferential tool, called Kullback-Leibler projection, to identify differentially methylated regions.

For the DNA methylation data, we observed that short-read counts $N(i,s,g)$ varied greatly with position $i$, and thus simply dividing the range of $Y_{i,s,g}$ can be misleading. Alternatively we first calculated the ratio of methylated read counts to total read counts at each position (these ratios are actually empirical estimates of the success probabilities along the genome). Then we divided the range of the ratios into disjoint intervals and followed the other steps in the adaptive basis selection algorithm. The size of the effective model space was $n^* = 100$ with $K = 10$ and $n_k = 10$, $k = 1, \ldots, K$. The CPU running time was about

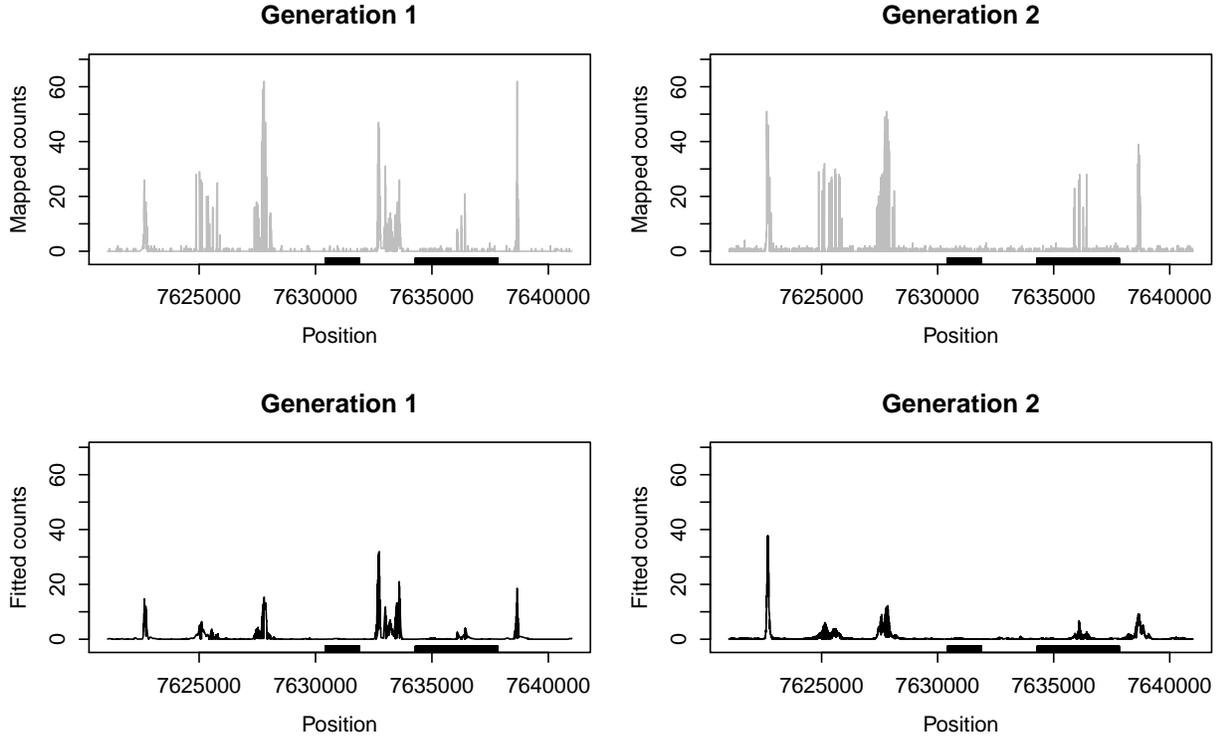

Figure 6: Mapped methylated read counts and fitted methylation level for a whole genome bisulfite sequencing data of *Arabidopsis thaliana*. The grey lines in top panels are the mapped methylation read counts for four strains of two generations. The black lines in bottom panels are the fitted methylation levels. The thick bars in x-axes are location of genes AT2G17540 (left) and AT2G17550 (right)

ten minutes in a computer with an Intel Xeon 2.90 GHz processor with 64GB of DDR3 RAM.

To identify differentially methylated regions between generations, we used Kullback-Leibler projection (Gu (2004)) for the "testing" of components which involve generation variable $g$, say $\eta_2(g)$ and $\eta_{12}(i,g)$. Generally, Kullback-Leibler projection helps assess the plausibility of the null hypothesis that $\eta$ belongs to a smaller space $\mathcal{H}^* \subset \mathcal{H}$. Thus, let $\hat{\eta}$ be the minimizer of penalized likelihood (2.1) in $\mathcal{H}$. Let $\tilde{\eta}$ be the Kullback-Leibler projection of $\hat{\eta}$ in a smaller space $\mathcal{H}^* \subset \mathcal{H}$, the minimizer of $\text{KL}(\hat{\eta}, \eta)$ for $\eta \in \mathcal{H}^*$. Let $\eta_C$ be the constant model as in (5.3). Based on an equality which holds for exponential family regression with canonical links (Gu (2004, Section 3.2)),

$$\text{KL}(\hat{\eta}, \eta_C) = \text{KL}(\hat{\eta}, \tilde{\eta}) + \text{KL}(\tilde{\eta}, \eta_C),$$

one can calculate the ratio $\rho = \text{KL}(\hat{\eta}, \tilde{\eta})/\text{KL}(\hat{\eta}, \eta_C)$ which quantifies how much of the struc-ture of $\hat{\eta}$ is lost by restricting within $\mathcal{H}^*$. A small $\rho$ indicates an adequacy of $\mathcal{H}^*$. For example, Gu (2004) suggests a reasonable threshold for $\rho$ could be in the range from 0.02 to 0.03.

An identified differentially methylated region (DMR) is plotted in Figure 6. This DMR is in *Arabidopsis* chromosome 2 ranging from genome position 7621000 to 7641000, which is in the intergenic region between gene AT2G17540 and gene AT2G17550 (TON1 RECRUITING MOTIF 26, TRM26). The Kullback-Leibler projection ratio for the model that contains only $\eta_1(i)$ is 0.11; the ratio for the model that contains both $\eta_1(i)$ and $\eta_2(g)$ is 0.08. Compared with a threshold 0.03, these ratios suggest that both $\eta_2(g)$ and $\eta_{12}(i,g)$ in (5.3) should be included. That is, the methylation levels among this region are differentiated between generations. Since this DMR is in the intergenic region of the TON1 gene, it is likely to be a partner of the TON1 gene. It was reported that in *Arabidopsis thaliana*, the TON1 proteins have a key regulatory role in microtubule organization at the cortex (Drevensek et al. (2012)). Thus, the identified DMR is likely to concertedly work with the TON1 proteins to regulate microtubule organization.

# 6 Discussion

Proper modeling of second-generation sequencing data plays an important role in navigating biological discovery. We develop an effective approximation of exponential family smoothing spline via adaptive basis selection for nonparametric modeling of second-generation sequencing data. With this, we constructed a lower-dimensional effective model space, in which exponential family smoothing spline models are estimated. We established the asymptotic convergence rate of smoothing splines via adaptive basis selection. More scalable computation make exponential family smoothing spline models via adaptive basis selection an appealing method for ultra-large sample sequencing data. We demonstrated its good performance in both simulated studies and data examples.

# A Supplementary Material

In this supplementary material, we further present some properties of our proposed estimator, and provide the proofs for related lemmas and theorems. Lastly, we show the derivation of generalized approximate cross-validation for choosing the tuning parameter.

## A.1 Properties of ABS estimator

Since our basis selection algorithm involves the response variable, the standard argument for the asymptotic analysis of smoothing splines does not apply. We first present some theoretical properties which shed light on how the adaptive basis sampling algorithm works and facilitate our asymptotic analysis.

Consider the estimation of $\mathrm{E}\{\psi(X,Y)\}$ based on $n$ i.i.d. observations $\{(x_i, y_i)\}_{i=1}^n$, where $\psi(x,y) \in \mathcal{L}^2(\mathcal{X}, \mathcal{Y})$ is a generic multivariate function. Two notations are introduced as the standard sample mean estimator and a mean estimator based on a subset of samples which is adaptively selected by our proposed method,

$$\mathrm{E}_n(\psi) = \frac{1}{n} \sum_{i=1}^n \psi(x_i, y_i), \tag{A.1}$$

$$\mathbb{E}_n^*(\psi) = \sum_{k=1}^K \frac{|S_k|}{n} \left\{ \frac{1}{n_k} \sum_{j=1}^{n_k} \psi(x_j^{*(k)}, y_j^{*(k)}) \right\}. \tag{A.2}$$

The following lemma shows the new estimator based on a subsample provides a good approximation to that based on all observations.

**Lemma 1.** *Suppose $n_k = n^*/K$, for $k = 1, \ldots, K$, then under the adaptive basis sampling scheme, the conditional variance of $\mathbb{E}_n^*(\psi)$ is bounded*

$$\mathrm{var}\{\mathbb{E}_n^*(\psi) | \{(x_i, y_i)\}_{i=1}^n\} \leqslant \frac{K}{n^*} \frac{1}{n} \sum_{i=1}^n \psi^2(x_i, y_i) \tag{A.3}$$

*and*

$$\mathrm{E}\{\mathbb{E}_n^*(\psi) - \mathrm{E}_n(\psi)\}^2 \leqslant \frac{K}{n^*} \mathrm{E}(\psi^2). \tag{A.4}$$

This lemma implies $\mathbb{E}_n^*(\psi) - \mathrm{E}_n(\psi)$ converges to zero in probability if $n^* \to \infty$ for $\psi$ with $E\{\psi^2(X,Y)\} < \infty$. In other words, the subsample estimator, $\mathbb{E}_n^*(\psi)$, is a good surrogate of the usual estimator $\mathrm{E}_n(\psi)$.

To understand the behavior of $\hat{\eta}_A$, the smoothing spline estimator computed using the adaptive basis selection algorithm, we refer to two important properties of the effective model space $\mathcal{H}_E$.

**Lemma 2.** *For any function outside the effective model space, its evaluations at selected samples $\{x_j^*\}_{j=1}^{n^*}$ are all zeros, i.e. for $h \in \mathcal{H} \ominus \mathcal{H}_E$,*

$$h(x_j^*) = 0, \qquad j = 1, \ldots, n^*.$$

**Lemma 3.** *Under Condition 1, 2, and 4, as $\lambda \to 0$ and $n^* \lambda^{2/r} \to \infty$, if unction $h$ is not in the effective model space, i.e., $h \in \mathcal{H} \ominus \mathcal{H}_E$, we have*

$$V(h) = o_p\{\lambda J(h)\}.$$

## A.2 Proofs

**Proof of Lemma 1** For each $k$, $1 \leq k \leq K$, $\{x_j^{*(k)}\}_{j=1}^{n_k}$ is a random draw from the $k$-th slice $S_k$. Thus, for $j = 1, \ldots, n_k$, the conditional mean of $\psi(x_j^{*(k)}, y_j^{*(k)})$ given the data is

$$\mathrm{E}\{\psi(x_j^{*(k)}, y_j^{*(k)}) | \{(x_i, y_i)\}_{i=1}^n\} = \frac{1}{|S_k|} \sum_{i=1}^n \psi(x_i, y_i) \mathbf{1}(y_i \in S_k). \tag{A.5}$$

It follows that the conditional mean of $\mathbb{E}_n^*(\psi)$ given the data is

$$\mathrm{E}\{\mathbb{E}_n^*(\psi) | \{(x_i, y_i)\}_{i=1}^n\}$$

$$= \mathrm{E}\left[\sum_{k=1}^K \frac{|S_k|}{n} \left\{\frac{1}{n_k} \sum_{j=1}^{n_k} \psi(x_j^{*(k)}, y_j^{*(k)})\right\} \bigg| \{(x_i, y_i)\}_{i=1}^n\right]$$

$$= \frac{1}{n} \sum_{k=1}^K \sum_{i=1}^n \psi(x_i, y_i) \mathbf{1}(y_i \in S_k) = \frac{1}{n} \sum_{i=1}^n \psi(x_i, y_i) = \mathrm{E}_n(\psi).$$

Hence $\mathbb{E}_n^*(\psi)$ and $\mathrm{E}_n(\psi)$ have the same mean value, $\mathrm{E}(\psi)$.

In the $k$-th slice, for $j = 1, \ldots, n_k$, the conditional variance of $\psi(x_j^{*(k)}, y_j^{*(k)})$ given the data is bounded by its second order conditional moment whose explicit form can be obtained by replacing $\psi$ by $\psi^2$ in (A.5), i.e.

$$\mathrm{var}\{\psi(x_j^{*(k)}, y_j^{*(k)}) | \{(x_i, y_i)\}_{i=1}^n\} \leqslant \mathrm{E}\{\psi^2(x_j^{*(k)}, y_j^{*(k)}) | \{(x_i, y_i)\}_{i=1}^n\} = \frac{1}{|S_k|} \sum_{i=1}^n \psi^2(x_i, y_i) \mathbf{1}(y_i \in S_k). \tag{A.6}$$

Noticing that samples from the same slice and from different slices are mutually independent, we obtain that

$$\mathrm{var}\{\mathbb{E}_n^*(\psi) | \{(x_i, y_i)\}_{i=1}^n\}$$

$$= \mathrm{var}\left[\sum_{k=1}^K \frac{|S_k|}{n} \left\{\frac{1}{n_k} \sum_{j=1}^{n_k} \psi(x_j^{*(k)}, y_j^{*(k)})\right\} \bigg| \{(x_i, y_i)\}_{i=1}^n\right]$$

$$= \sum_{k=1}^K \frac{|S_k|^2}{n^2} \frac{1}{n_k} \mathrm{var}\{\psi(x_j^{*(k)}, y_j^{*(k)}) | \{(x_i, y_i)\}_{i=1}^n\}.$$

The right hand side of the above has an upper bound due to (A.6)

$$\mathrm{var}\{\mathbb{E}_n^*(\psi) | \{(x_i, y_i)\}_{i=1}^n\} \leqslant \sum_{k=1}^K \frac{|S_k|}{n^2} \frac{1}{n_k} \sum_{i=1}^n \psi^2(x_i, y_i) \mathbf{1}(y_i \in S_k),$$

which in turn is upper bounded by

$$\sum_{k=1}^K \frac{1}{n} \frac{1}{n^*/K} \sum_{i=1}^n \psi^2(x_i, y_i) \mathbf{1}(y_i \in S_k) = \frac{K}{n^*} \frac{1}{n} \sum_{i=1}^n \psi^2(x_i, y_i)$$

with the fact that $n_k = n^*/K$ and $|S_k|/n \leqslant 1$. We thus have proved (A.3).

The condition mean of $\mathbb{E}_n^*(\psi)$ given the data has been proved to be $\mathrm{E}_n(\psi)$. Recall the definition of conditional variance, we have

$$\mathrm{var}\{\mathbb{E}_n^*(\psi)|\{(x_i, y_i)\}_{i=1}^n\} = \mathrm{E}[\{\mathbb{E}_n^*(\psi) - \mathrm{E}_n(\psi)\}^2|\{(x_i, y_i)\}_{i=1}^n].$$

We obtain (A.4) immediately by taking expectation on both sides of the above, i.e.

$$\mathrm{E}\{\mathbb{E}_n^*(\psi) - \mathrm{E}_n(\psi)\}^2 = \mathrm{E}[\mathrm{var}\{\mathbb{E}_n^*(\psi)|\{(x_i, y_i)\}_{i=1}^n\}] \leqslant \frac{K}{n^*} \mathrm{E}(\psi^2).$$

Before proving the main result, we first present two useful lemmas in Gu (2013).

**Lemma 4.** *Under Condition 2, as $\lambda \to 0$, one has*

$$\sum_\nu \frac{1}{1 + \lambda \rho_\nu} = O(\lambda^{-1/r}).$$

This is part of Lemma 9.1 in Gu (2013).

**Lemma 5.** *Under Condition 1, 2 and 4, as $\lambda \to 0$ and $n\lambda^{2/r} \to \infty$,*

$$\frac{1}{n} \sum_{i=1}^n g(x_i)h(x_i)w(\eta_0(x_i); y_i) = V(g, h) + o_p(\{(V + \lambda J)(g)(V + \lambda J)(h)\}^{1/2})$$

*for all $g$ and $h$ in $\mathcal{H}$.*

This is Lemma 9.16 in Gu (2013).

**Proof of Lemma 2** See the supplementary material of Ma et al. (2015).

**Proof of Lemma 3** By Lemma 2, given the selected samples $\{x_j^*\}_{j=1}^{n^*}$, for any $h \in \mathcal{H} \ominus \mathcal{H}_E$, we have

$$h(x_j^*) = 0 \qquad j = 1, \ldots, n^*.$$

Note that $\{x_j^*\}_{j=1}^{n^*}$ is the collection of $\{x_j^{*(k)}\}_{j=1}^{n_k}$ from $k = 1, \ldots, K$ slices, hence

$$\mathbb{E}_n^*\{h^2(X)w(\eta_0(X); Y)\} = \sum_{k=1}^K \frac{|S_k|}{n}\left\{\frac{1}{n_k}\sum_{j=1}^{n_k} h^2(x_j^{*(k)})w(\eta_0(x_j^{*(k)}); y_j^{*(k)})\right\} = 0.$$

It follows that

$$V(h) = \int_\mathcal{X} h^2(x)v_{\eta_0}(x)f_X(x)\,dx = \mathrm{E}\{h(X)^2 v_{\eta_0}(X)\} - \mathbb{E}_n^*\{h^2(X)w(\eta_0(X); Y))\}. \tag{A.7}$$

By Condition 1, there exist a collection of functions $\phi_\nu \in \mathcal{H}$ and a sequence of nonnegative $\rho_\nu$ such that $V$ and $J$ are simultaneously diagonalized, i.e., $V(\phi_\nu, \phi_\mu) = \delta_{\nu\mu}$ and $J(\phi_\nu, \phi_\mu) = \rho_\nu \delta_{\nu\mu}$. Use $\phi_\nu$'s as basis functions and expand $h$ as $h = \sum_\nu h_\nu \phi_\nu$, where $h_\nu = V(h, \phi_\nu)$. Then, (A.7) can be written as

$$V(h) = \mathrm{E}\left\{\left(\sum_\nu h_\nu \phi_\nu(X)\right)^2 v_{\eta_0}(X)\right\} - \mathbb{E}_n^*\left\{\left(\sum_\nu h_\nu \phi_\nu(X)\right)^2 w(\eta_0(X); Y)\right\}.$$

Due to the fact that $\mathrm{E}(\cdot)$ and $\mathbb{E}_n^*(\cdot)$ are both linear operators, we have

$$V(h) = \sum_\nu \sum_\mu h_\nu h_\mu \big[\mathrm{E}\{\phi_\nu(X)\phi_\mu(X)v_{\eta_0}(X)\} - \mathbb{E}_n^*\{\phi_\nu(X)\phi_\mu(X)w(\eta_0(X); Y)\}\big].$$

Applying the Cauchy-Schwarz inequality to obtain

$$V(h) \leqslant I^{1/2} \cdot \left\{\sum_\nu \sum_\mu h_\nu^2 h_\mu^2 (1 + \lambda\rho_\nu)(1 + \lambda\rho_\mu)\right\}^{1/2} \tag{A.8}$$

$$= I^{1/2} \cdot \sum_\nu h_\nu^2 (1 + \lambda\rho_\nu) \tag{A.9}$$

where

$$I = \sum_\nu \sum_\mu \frac{1}{1 + \lambda\rho_\nu} \frac{1}{1 + \lambda\rho_\mu} \big[\mathrm{E}\{\phi_\nu(X)\phi_\mu(X)v_{\eta_0}(X)\} - \mathbb{E}_n^*\{\phi_\nu(X)\phi_\mu(X)w(\eta_0(X); Y)\}\big]^2. \tag{A.10}$$

Since $\phi_\nu$'s simultaneously diagonalize $V$ and $J$,

$$\sum_\nu h_\nu^2 (1 + \lambda\rho_\nu) = (V + \lambda J)(h). \tag{A.11}$$

In light of (A.8), to bound $V(h)$, we need to investigate the magnitude of $I$ whose expression is given in (A.10).

First, by inserting

$$\mathrm{E}_n\{\phi_\nu(X)\phi_\mu(X)w(\eta_0(X); Y)\} = \frac{1}{n}\sum_{i=1}^n \phi_\nu(x_i)\phi_\mu(x_i)w(\eta_0(x_i); y_i)$$

into the squared term in (A.10) and applying the inequality $(a + b)^2 \leqslant 2a^2 + 2b^2$, we obtain

$$I \leqslant 2\sum_\nu \sum_\mu \frac{1}{1+\lambda\rho_\nu}\frac{1}{1+\lambda\rho_\mu}\big[\mathrm{E}\{\phi_\nu(X)\phi_\mu(X)v_{\eta_0}(X)\} - \mathrm{E}_n\{\phi_\nu(X)\phi_\mu(X)w(\eta_0(X); Y)\}\big]^2$$

$$+ 2\sum_\nu \sum_\mu \frac{1}{1+\lambda\rho_\nu}\frac{1}{1+\lambda\rho_\mu}\big[\mathrm{E}_n\{\phi_\nu(X)\phi_\mu(X)w(\eta_0(X); Y)\} - \mathbb{E}_n^*\{\phi_\nu(X)\phi_\mu(X)w(\eta_0(X); Y)\}\big]^2$$

$$\triangleq 2I_1 + 2I_2.$$

Next, we examine the magnitudes of $I_1$ and $I_2$ one by one.

*Order of $I_1$.* Recall that $E\{w(\eta_0(x); y)\} = v_{\eta_0}(x)$, then

$$E\big[E_n\{\phi_\nu(X)\phi_\mu(X)w(\eta_0(X); Y)\}\big] = E\{\phi_\nu(X)\phi_\mu(X)v_{\eta_0}(X)\}$$

and

$$\text{var}\big[E_n\{\phi_\nu(X)\phi_\mu(X)w(\eta_0(X); Y)\}\big] = \frac{1}{n}\text{var}\{\phi_\nu(X)\phi_\mu(X)w(\eta_0(X); Y)\}.$$

Therefore, the expectation of $I_1$ is

$$E\, I_1 = \sum_\nu \sum_\mu \frac{1}{1+\lambda\rho_\nu}\frac{1}{1+\lambda\rho_\mu} E\big[E\{\phi_\nu(X)\phi_\mu(X)v_{\eta_0}(X)\} - E_n\{\phi_\nu(X)\phi_\mu(X)w(\eta_0(X); Y)\}\big]^2$$

$$= \sum_\nu \sum_\mu \frac{1}{1+\lambda\rho_\nu}\frac{1}{1+\lambda\rho_\mu} \frac{1}{n}\text{var}\{\phi_\nu(X)\phi_\mu(X)w(\eta_0(X); Y)\}.$$

By Condition 4, $\text{var}\{\phi_\nu(X)\phi_\mu(X)w(\eta_0(X); Y)\} \leqslant c_3$ for some constant $c_3 < \infty$. Hence, by Lemma 4,

$$E\, I_1 \leqslant \frac{c_3}{n}\left(\sum_\nu \frac{1}{1+\lambda\rho_\nu}\right)^2 = O(n^{-1}\lambda^{-2/r}). \tag{A.12}$$

*Order of $I_2$.* The expectation of $I_2$ is

$$E\, I_2 = \sum_\nu \sum_\mu \frac{1}{1+\lambda\rho_\nu}\frac{1}{1+\lambda\rho_\mu} E\big[E_n\{\phi_\nu(X)\phi_\mu(X)w(\eta_0(X); Y)\} - \mathbb{E}_n^*\{\phi_\nu(X)\phi_\mu(X)w(\eta_0(X); Y)\big]^2.$$

As in Lemma 1, we assume $n_k = n^*/K$ for all $k$ and substitute $\psi(x, y)$ by $\phi_\nu(x)\phi_\mu(x)w(\eta_0(x); y)$ in (A.4) to obtain

$$E\big[E_n\{\phi_\nu(X)\phi_\mu(X)w(\eta_0(X); Y)\} - \mathbb{E}_n^*\{\phi_\nu(X)\phi_\mu(X)w(\eta_0(X); Y)\big]^2$$
$$\leqslant \frac{K}{n^*}E\{\phi_\nu^2(X)\phi_\mu^2(X)w^2(\eta_0(X); Y)\}$$
$$\leqslant \frac{K}{n^*}(c_3+1),$$

where the constant $c_3$ is the bound of $\text{var}\{\phi_\nu(X)\phi_\mu(X)w(\eta_0(X); Y)\}$ in Condition 4. Again, by Lemma 4,

$$E\, I_2 \leqslant \frac{K(c_3+1)}{n^*}\left(\sum_\nu \frac{1}{1+\lambda\rho_\nu}\right)^2 = O(n^{*-1}\lambda^{-2/r}). \tag{A.13}$$

Putting (A.12) and (A.13) together and noticing $n^* \ll n$, we obtain

$$E\, I \leqslant 2\, E\, I_1 + 2\, E\, I_2 = O(n^{*-1}\lambda^{-2/r}) + O(n^{-1}\lambda^{-2/r}) = O(n^{*-1}\lambda^{-2/r}).$$

Therefore $I = O_p(n^{*-1}\lambda^{-2/r})$ and $V(h) \leqslant (V + \lambda J)(h) \cdot O_p(n^{*-1/2}\lambda^{-1/r})$. The desired result follows from the fact $n^{*-1/2}\lambda^{-1/r} \to 0$.

**Proof of Theorem 1 in the main paper** By the representer theorem, $\hat{\eta}$, the minimizer of (2.1) in the main paper, has an explicit form as in (2.2) of the main paper. Given the effective model space $\mathcal{H}_E$, let $\hat{\eta}_E$ be the projection of $\hat{\eta}$ to $\mathcal{H}_E$ relative to the reproducing kernel Hilbert space inner product. The proposed estimator $\hat{\eta}_A$ uses basis functions from $\mathcal{H}_E$ while $\hat{\eta}$ uses the full basis from $\mathcal{H}$.

According to Theorem 9.17 in Gu (2013), $\hat{\eta}$ converges to the true function $\eta_0$ with certain rate. Notice that

$$\hat{\eta}_A - \eta_0 = (\hat{\eta}_A - \hat{\eta}_E) + (\hat{\eta}_E - \hat{\eta}) + (\hat{\eta} - \eta_0).$$

It suffices to show that both $\hat{\eta}_E - \hat{\eta}$ and $\hat{\eta}_A - \hat{\eta}_E$ converge to zero at the same or a faster rate. We achieve this in two steps.

*Step 1.* We show that $\hat{\eta}_E$ converges to $\eta_0$ with the same rate as $\hat{\eta}$. To this end, note that $\hat{\eta} - \hat{\eta}_E \in \mathcal{H} \ominus \mathcal{H}_E \subseteq \mathcal{H}_J$ and $\hat{\eta} \in \mathcal{H}_E$, therefore $J(\hat{\eta} - \hat{\eta}_E, \hat{\eta}_E) = 0$.

For any functions $g, h \in \mathcal{H}$, define

$$A_{g,h}(\alpha) = \frac{1}{n}\sum_{i=1}^{n} l\{(g + \alpha h)(x_i); y_i\} + \frac{\lambda}{2}J(g + \alpha h).$$

It can be easily shown that

$$\left.\frac{dA_{g,h}(\alpha)}{d\alpha}\right|_{\alpha=0} = \frac{1}{n}\sum_{i=1}^{n} u(g(x_i); y_i)h(x_i) + \lambda J(g, h). \tag{A.14}$$

Since $\hat{\eta}$ is the minimizer of (2) in the main paper over $\mathcal{H}$, $A_{g,h}(\alpha)$ reaches its minimum at $\alpha = 0$ when $g = \hat{\eta}$ and $h = \hat{\eta} - \hat{\eta}_E$. Thus, for this choice of $g$ and $h$, the derivative in (A.14) is zero. It follows that

$$\lambda J(\hat{\eta}, \hat{\eta} - \hat{\eta}_E) = -\frac{1}{n}\sum_{i=1}^{n} u(\hat{\eta}(x_i); y_i)\{\hat{\eta}(x_i) - \hat{\eta}_E(x_i)\}. \tag{A.15}$$

The fact that $J(\hat{\eta} - \hat{\eta}_E, \hat{\eta}_E) = 0$ implies $J(\hat{\eta} - \hat{\eta}_E)$ is equal to $J(\hat{\eta}, \hat{\eta} - \hat{\eta}_E)$. Thus

$$\lambda J(\hat{\eta} - \hat{\eta}_E) = -\frac{1}{n}\sum_{i=1}^{n} u(\hat{\eta}(x_i); y_i)\{\hat{\eta}(x_i) - \hat{\eta}_E(x_i)\} \triangleq S_1 + S_2, \tag{A.16}$$

where

$$S_1 = -\frac{1}{n}\sum_{i=1}^{n}\{u(\hat{\eta}(x_i); y_i) - u(\eta_0(x_i); y_i)\}\{\hat{\eta}(x_i) - \hat{\eta}_E(x_i)\},$$

$$S_2 = -\frac{1}{n}\sum_{i=1}^{n} u(\eta_0(x_i); y_i)\{\hat{\eta}(x_i) - \hat{\eta}_E(x_i)\}.$$

We next study the orders of the two terms $S_1$ and $S_2$ under Conditions 1, 2 and 4, and $\lambda \to 0$, $n\lambda^{2/r} \to \infty$.

For $S_1$, since $u(\eta(x), y)$ is differentiable with respect to $\eta(x)$, it follows by the mean value theorem and Condition 3 that there exists a constant $\gamma \in [c_1, c_2]$ such that

$$S_1 = -\frac{\gamma}{n} \sum_{i=1}^n w(\eta_0(x_i); y_i)\{\hat{\eta}(x_i) - \eta_0(x_i)\}\{\hat{\eta}(x_i) - \hat{\eta}_E(x_i)\}.$$

Applying Lemma 5 to the right hand side of the above, we have

$$|S_1| = \gamma V(\hat{\eta} - \eta_0, \hat{\eta} - \hat{\eta}_E) + \{(V + \lambda J)(\hat{\eta} - \eta_0)(V + \lambda J)(\hat{\eta} - \hat{\eta}_E)\}^{1/2} o_p(1)$$
$$= \{(V + \lambda J)(\hat{\eta} - \eta_0)(V + \lambda J)(\hat{\eta} - \hat{\eta}_E)\}^{1/2} O_p(1)$$

For $S_2$, recall $\phi_\nu \in \mathcal{H}$ are eigenfunctions which simultaneously diagonalize $V$ and $J$ such that $V(\phi_\nu, \phi_\mu) = \delta_{\nu\mu}$ and $J(\phi_\nu, \phi_\mu) = \rho_\nu \delta_{\nu\mu}$. Write $\hat{\eta} - \hat{\eta}_E = \sum_\nu (\hat{\eta} - \hat{\eta}_E)_\nu \phi_\nu$, where $(\hat{\eta} - \hat{\eta}_E)_\nu = V(\hat{\eta} - \hat{\eta}_E, \phi_\nu)$. Plugging it in $S_2$ and applying Cauchy-Schwarz inequality, we have

$$|S_2| = \left|\sum_\nu (\hat{\eta} - \hat{\eta}_E)_\nu \left\{\frac{1}{n}\sum_{i=1}^n u(\eta_0(x_i); y_i)\phi_\nu(x_i)\right\}\right|$$
$$\leq \left\{\sum_\nu \frac{\beta_\nu^2}{1+\lambda\rho_\nu}\right\}^{1/2} \left\{\sum_\nu (\hat{\eta} - \hat{\eta}_E)_\nu^2(1+\lambda\rho_\nu)\right\}^{1/2}$$

where $\beta_\nu = \frac{1}{n}\sum_{i=1}^n u(\eta_0(x_i); y_i)\phi_\nu(x_i)$ possesses properties $\mathrm{E}(\beta_\nu) = 0$ and $\mathrm{var}(\beta_\nu) = \sigma^2/n$. In fact

$$\mathrm{E}(\beta_\nu) = \mathrm{E}\{u(\eta_0(X); Y)\phi_\nu(X)\} = \mathrm{E}_X\big[\mathrm{E}\{u(\eta_0(X); Y)|X\}\phi_\nu(X)\big] = 0$$

and

$$\mathrm{E}(\beta_\nu^2) = \frac{1}{n}\mathrm{E}\{u^2(\eta_0(X); Y)\phi_\nu^2(X)\} = \frac{1}{n}\mathrm{E}_X\big[\mathrm{E}\{u^2(\eta_0(X); Y)|X\}\phi_\nu^2(X)\}\big]$$
$$= \frac{\sigma^2}{n}\mathrm{E}_X\{v_{\eta_0}(X)\phi_\nu^2(X)\} = \frac{\sigma^2}{n}V(\phi_\nu) = \frac{\sigma^2}{n}.$$

Furthermore, by Lemma 4,

$$\mathrm{E}\left\{\sum_\nu \frac{\beta_\nu^2}{1+\lambda\rho_\nu}\right\} = \frac{\sigma^2}{n}\sum_\nu \frac{1}{1+\lambda\rho_\nu} = O(n^{-1}\lambda^{-1/r}). \quad (A.17)$$

and it can be shown by a similar argument as in (A.11) that

$$\sum_\nu (\hat{\eta} - \hat{\eta}_E)_\nu^2(1+\lambda\rho_\nu) = (V + \lambda J)(\hat{\eta} - \hat{\eta}_E). \quad (A.18)$$

Combining (A.17) and (A.18), we obtain

$$S_2 \leqslant \{(V + \lambda J)(\hat{\eta} - \hat{\eta}_E)\}^{1/2} O_p(n^{-1/2}\lambda^{-1/(2r)}).$$

Now we are ready to determine the order of $(V + \lambda J)(\hat{\eta} - \hat{\eta}_E)$. By Lemma 3, $V(\hat{\eta} - \hat{\eta}_E)$ is dominated by $\lambda J(\hat{\eta} - \hat{\eta}_E)$ since $\hat{\eta} - \hat{\eta}_E \in \mathcal{H} \ominus \mathcal{H}_E$. Thus, $(V + \lambda J)(\hat{\eta} - \hat{\eta}_E)$ converges to zero at the same order as $\lambda J(\hat{\eta} - \hat{\eta}_E)$. Therefore, it follows (A.16) that

$$\begin{aligned}(V + \lambda J)(\hat{\eta} - \hat{\eta}_E) &\asymp \lambda J(\hat{\eta} - \hat{\eta}_E) = S_1 + S_2 \\ &\leqslant \{(V + \lambda J)(\hat{\eta} - \eta_0)(V + \lambda J)(\hat{\eta} - \hat{\eta}_E)\}^{1/2} O_p(1) \\ &\quad + \{(V + \lambda J)(\hat{\eta} - \hat{\eta}_E)\}^{1/2} O_p(n^{-1/2}\lambda^{-1/(2r)}).\end{aligned}$$

After canceling out $\{(V + \lambda J)(\hat{\eta} - \hat{\eta}_E)\}^{1/2}$ and taking squares on both sides, we obtain

$$\begin{aligned}(V + \lambda J)(\hat{\eta} - \hat{\eta}_E) &\leqslant (V + \lambda J)(\hat{\eta} - \eta_0) O_p(1) + O_p(n^{-1}\lambda^{-1/r}) \\ &\asymp (V + \lambda J)(\hat{\eta} - \eta_0) \\ &= O_p(n^{-1}\lambda^{-1/r} + \lambda^p).\end{aligned}$$

*Step 2.* We show that $\hat{\eta}_A$, the smoothing spline estimator via adaptive sampling scheme, converges to $\eta_0$ with the same convergence rate as $\hat{\eta}_E$.

Since $\hat{\eta}$ is the minimizer of (2.2) in the main paper over $\mathcal{H}$, $A_{g,h}(\alpha)$ reaches its minimum at $\alpha = 0$ when $g = \hat{\eta}$ and $h = \hat{\eta}_A - \hat{\eta}_E$. Arguing as in the proof of (A.15), we have

$$\lambda J(\hat{\eta}, \hat{\eta}_A - \hat{\eta}_E) = -\frac{1}{n} \sum_{i=1}^n u(\hat{\eta}(x_i); y_i)\{\hat{\eta}_A(x_i) - \hat{\eta}_E(x_i)\}. \tag{A.19}$$

Since $\hat{\eta}_A$ is also the minimizer of (2.2) in the main paper over $\mathcal{H}_E$, $A_{g,h}(\alpha)$ reaches its minimum at $\alpha = 0$ when $g = \hat{\eta}_A$ and $h = \hat{\eta}_A - \hat{\eta}_E$. Thus, similar to the previous result, we have

$$\lambda J(\hat{\eta}_A, \hat{\eta}_A - \hat{\eta}_E) = -\frac{1}{n} \sum_{i=1}^n u(\hat{\eta}_A(x_i); y_i)\{\hat{\eta}_A(x_i) - \hat{\eta}_E(x_i)\}. \tag{A.20}$$

We subtract (A.19) from (A.20) to obtain

$$\lambda J(\hat{\eta}_A - \hat{\eta}, \hat{\eta}_A - \hat{\eta}_E) = \frac{1}{n} \sum_{i=1}^n \{u(\hat{\eta}(x_i); y_i) - u(\hat{\eta}_A(x_i); y_i)\}\{\hat{\eta}_A(x_i) - \hat{\eta}_E(x_i)\}.$$

Recall that $\hat{\eta}_E$ is the projection of $\hat{\eta}$ onto $\mathcal{H}_E$ and $\hat{\eta}_A - \hat{\eta}_E \in \mathcal{H}_E$, then $(\hat{\eta} - \hat{\eta}_E) \perp (\hat{\eta}_A - \hat{\eta}_E)$. Such orthogonality implies that $J(\hat{\eta} - \hat{\eta}_E, \hat{\eta}_A - \hat{\eta}_E) = 0$ and further

$$J(\hat{\eta}_A - \hat{\eta}_E) = J(\hat{\eta}_A - \hat{\eta}, \hat{\eta}_A - \hat{\eta}_E) + J(\hat{\eta} - \hat{\eta}_E, \hat{\eta}_A - \hat{\eta}_E) = J(\hat{\eta}_A - \hat{\eta}, \hat{\eta}_A - \hat{\eta}_E).$$

With this result, some algebra yields

$$\frac{1}{n}\sum_{i=1}^n \{u(\hat{\eta}_A(x_i); y_i) - u(\hat{\eta}_E(x_i); y_i)\}\{\hat{\eta}_A(x_i) - \hat{\eta}_E(x_i)\} + \lambda J(\hat{\eta}_A - \hat{\eta}_E) \quad (A.21)$$

$$= \frac{1}{n}\sum_{i=1}^n \{u(\hat{\eta}(x_i); y_i) - u(\hat{\eta}_E(x_i); y_i)\}\{\hat{\eta}_A(x_i) - \hat{\eta}_E(x_i)\} \quad (A.22)$$

By the mean value theorem, Condition 3 and Lemma 5, there exists a constant $\zeta \in [c_1, c_2]$ such that the left hand side of (A.21) equals

$$\zeta V(\hat{\eta}_A - \hat{\eta}_E) + o_p\{(V + \lambda J)(\hat{\eta}_A - \hat{\eta}_E)\} + \lambda J(\hat{\eta}_A - \hat{\eta}_E) = (V + \lambda J)(\hat{\eta}_A - \hat{\eta}_E)\{1 + o_p(1)\}.$$

Similarly the right hand side of (A.21) is bounded by

$$\{(V + \lambda J)(\hat{\eta} - \hat{\eta}_E)(V + \lambda J)(\hat{\eta}_A - \hat{\eta}_E)\}^{1/2} O_p(1).$$

Combining the above two results, we obtain that

$$(V + \lambda J)(\hat{\eta}_A - \hat{\eta}_E)\{1 + o_p(1)\} = \{(V + \lambda J)(\hat{\eta} - \hat{\eta}_E)(V + \lambda J)(\hat{\eta}_A - \hat{\eta}_E)\}^{1/2} O_p(1).$$

Canceling out a term from both sides to obtain

$$(V + \lambda J)(\hat{\eta}_A - \hat{\eta}_E) \asymp (V + \lambda J)(\hat{\eta} - \hat{\eta}_E) = O_p(n^{-1}\lambda^{-1/r} + \lambda^p). \quad (A.23)$$

Putting results from Step 1 and 2 together, we conclude the proof with the convergence rate

$$(V + \lambda J)(\hat{\eta}_A - \eta_0) = O_p(n^{-1}\lambda^{-1/r} + \lambda^p).$$

## A.3 Derivation of generalized approximate cross-validation

The minimizer of (2.6) in the main paper satisfies the normal equation

$$\begin{pmatrix} S_w^T S_w & S_w^T R_w \\ R_w^T R_w & R_w^T R_w + (n\lambda)Q \end{pmatrix} \begin{pmatrix} \mathbf{d} \\ \mathbf{c} \end{pmatrix} = \begin{pmatrix} S_w^T \tilde{\mathbf{Y}}_w \\ R_w^T \tilde{\mathbf{Y}}_w \end{pmatrix}, \quad (A.24)$$

where $S_w = \tilde{W}^{1/2}S$, $R_w = \tilde{W}^{1/2}R$, and $\tilde{\mathbf{Y}}_w = \tilde{W}^{1/2}\tilde{\mathbf{Y}}$. The normal equation of (A.24) can be solved by the pivoted Cholesky decomposition followed by backward and forward substitutions (Kim and Gu (2004)). On the convergence of Newton iteration, the "fitted values" of $\hat{\mathbf{Y}}_w = S_w \mathbf{d} + R_w \mathbf{c}$ by minimizing (2.4) in the main paper can be written as $\hat{\mathbf{Y}}_w = A_w(\lambda)\tilde{\mathbf{Y}}_w$, where the smoothing matrix

$$A_w(\lambda) = (S_w, R_w) \begin{pmatrix} S_w^T S_w & S_w^T R_w \\ R_w^T R_w & R_w^T R_w + (n\lambda)Q \end{pmatrix}^+ \begin{pmatrix} S_w^T \\ R_w^T \end{pmatrix},$$

where $S_w = \tilde{W}^{1/2}S$, $R_w = \tilde{W}^{1/2}R$, $\tilde{\mathbf{Y}}_w = \tilde{W}^{1/2}\tilde{\mathbf{Y}}$, and $\mathbf{C}^+$ denotes the Moore-Penrose inverse of $\mathbf{C}$ satisfying $\mathbf{CC}^+\mathbf{C} = \mathbf{C}$, $\mathbf{C}^+\mathbf{CC}^+ = \mathbf{C}^+$, $(\mathbf{CC}^+)^T = \mathbf{CC}^+$ and $(\mathbf{C}^+\mathbf{C})^T = \mathbf{C}^+\mathbf{C}$.

A data-driven approach for the selection of the tuning parameter $\lambda$ (including $\theta$) is to choose $\lambda$ which minimizes the generalized approximate cross-validation score (Gu and Xiang (2001)), one version of which was derived by Gu and Xiang (2001) and is of the following form,

$$GACV(\lambda) = -\frac{1}{n}\sum_{i=1}^{n}\{Y_i\hat{\eta}_A(x_i) - b(\hat{\eta}_A(x_i))\} + \frac{\text{tr}(A_w\tilde{W}^{-1})}{n - \text{tr}A_w}\frac{1}{n}\sum_{i=1}^{n}Y_i(Y_i - \hat{\mu}(x_i)). \quad (A.25)$$

One may employ standard nonlinear optimization algorithms to minimize the generalized approximate cross-validation score. In particular, we use the modified Newton algorithm developed by Dennis and Schnabel (1996) to find the minimizer. $\hat{\eta}_A$ and $\hat{\mu}$ are evaluated at the minimizer of (2.1) in the main paper with fixed tuning parameters, and $A_w$ and $\tilde{W}$ are evaluated at the values given at the convergence of the Newton iterations.